\documentclass[reqno]{article}
\newcommand{\ds}{\displaystyle}
\newcommand{\dsf}{\ds\frac}

\newcommand{\beq}{\begin{equation}}
\newcommand{\eeq}{\end{equation}}

\usepackage{ae} 
\usepackage[T1]{fontenc}
\usepackage[ansinew]{inputenc}
\usepackage{amsmath}
\usepackage{graphicx}
\usepackage{color}
\usepackage[colorlinks]{hyperref}
\usepackage{multicol}
\begin{document}
\small

\begin{center}
\bf Flux instabilities in type-II superconductors
\end{center}
\begin{center}
N. A. Taylanov
\end{center}
\begin{center}
\emph{\footnotesize National University of Uzbekistan}
\end{center}
\begin{center}
 \textcolor{black}{Abstract}
\end{center}
\begin{center}
\mbox{\parbox{14cm}{\footnotesize The flux jump dynamics in the
flux flow regime of type II superconductors is investigated,
analytically. It is found that under some conditions flux jump
avalanche may occur in a superconductor sample, which takes into
account an inertial properties of the vortex matter.}}
\end{center}
{\bf Key words}: flux jumps, vortex mass, critical state, flux
flow.

\begin{multicols}{2}{
\vskip 0.5cm
\begin{center}
{\bf   Introduction}
\end{center}

As we know, the flux jumps results in a large-scale flux
avalanches in a superconductor and their origin are related to the
magnetothermal instabilities [1-5]. Thermomagnetic instability or
flux jump phenomena have been observed in conventional hard
superconductor, as well as in high-temperature superconductors,
recently [1-6]. The spatial and temporal development of this
instability depends on the sample geometry, temperature, external
magnetic field, its rate of change and orientation, initial and
boundary conditions, etc. The critical state instabilities result
in flux redistribution towards the equilibrium state and are
accompanied by a significant heat release, which often leads to
the superconductor-to-normal-transition. Recently, Chabanenko et
al. [6] have reported an interesting phenomenon in their
experiments - convergent oscillations of the magnetic flux arising
from flux jump avalanches [6-11]. The authors argued that the
observed oscillations due to flux avalanches can be interpreted as
a result of the existence of a definite value of the effective
vortex mass [12-21]. Thus, it is necessary to take into account
collective modes, i.e., the inertial properties of the vortices in
studying the dynamics of the flux avalanches. In the present work,
we study the dynamics of the magnetic flux avalanche, which takes
into account an inertial properties of the vortex matter.

\vskip 0.5cm
\begin{center}
{\bf  1. Formulation}
\end{center}
Bean [1] has proposed the critical state model which is
successfully used to describe magnetic properties of type II
superconductors. According to this model, the distribution of the
magnetic flux density $\vec B$ and the transport current density
$\vec j$ inside a superconductor is given by a solution of the
equation

\begin{equation}
rot\vec B=\mu_0\vec j.
\end{equation}
When the penetrated magnetic flux changes with time, an electric
field $\vec E(r, t)$ is generated inside the sample according to
Faraday's law

\begin{equation}
rot\vec E=\dsf{1}{c}\dsf{d\vec B}{dt}.
\end{equation}
In the flux flow regime the electric field $\vec E(r, t)$ induced
by the moving vortices is related with the local current density
$\vec j(r, t)$ by the nonlinear Ohm's law

\begin{equation}
\vec E=\vec v\vec B.
\end{equation}
To obtain qualitative results, we use a classical equation of
motion of a vortex, which it can derived by integrating over the
microscopic degrees of freedom, leaving only macroscopic forces
[21]. Thus, the equation of the vortex motion under the action of
the Lorentz, pinning, and viscosity forces can be presented as

\begin{equation}
m\dsf{dV}{dt}+\eta V+F_L+F_p=0.
\end{equation}
Here $m$ is the vortex mass per unit length, $\vec
F_L=\dsf{1}{c}\vec j\vec \Phi_0$ is the Lorentz force,  $\vec
F_p=\dsf{1}{c}\vec j_c\vec \Phi_0$,
$\eta=\dsf{\Phi_0H_{c2}}{c^2\rho_n}$ is the flux flow viscosity
coefficient, $\Phi_0=\pi hc/2e$ is the magnetic flux quantum,
$H_{c2}$ is the upper critical field of superconductor, $\rho_n$
is the normal state resistivity, $j_c$ is the critical current
density [4]. For simplicity we have neglected the Magnus force,
assuming that it is much smaller then the viscous force (for
example, for Nb see, [6]). In the absence of external currents and
fields, the Lorentz force results from currents associated with
vortices trapped in the sample.

\vskip 0.5cm
\begin{center}
{\bf 2. Basic equation}
\end{center}

In combining the relation (3) with Maxwell’s equation (2), we
obtain a nonlinear diffusion equation for the magnetic flux
induction $\vec B(r, t)$ in the following form
\begin{equation}
m\dsf{dV}{dt}+\eta V=-\dsf{1}{c}\Phi_0(j-j_c),
\end{equation}
\begin{equation}
\dsf{d\vec B}{dt}=\nabla [\vec v\vec B].
\end{equation}
The temperature distribution in superconductor is governed by the
heat conduction diffusion equation
\begin{equation}
\nu (T)\dsf{dT}{dt}=\nabla[\kappa(T)\nabla T]+\vec j\vec E,
\end{equation}
Here $\nu=\nu(T)$ and $\kappa=\kappa(T)$ are the specific heat and
thermal conductivity, respectively. The above equations should be
supplemented by a current-voltage characteristics of
superconductors, which has the form
$$
\vec j=j_{c}(T, \vec B, \vec E).
$$
In order to obtain analytical results of equations (5)-(7), we
suggest that $j_c$ is independent on magnetic field induction $B$
and use the Bean critical state model $j_c=j_c(B_e, T)$, i.e.,
$j_c(T)=j_0-a(T-T_0)$ [4];  where $B_e$ is the external applied
magnetic field induction, $a=j_0/(T_c-T_0)$, $T_0$ and $T_c$ are
the equilibrium and critical temperatures of the sample,
respectively, $j_0$ is the equilibrium current density. For the
sake of simplifying of the calculations, we perform our
calculations on the assumption of negligibly small heating and
assume that the temperature profile is a constant within the
across sample and thermal conductivity $\kappa$ and heat capacity
$\nu$ are independent on the temperature profile [5].

We study the evolution of the thermal and electromagnetic
penetration process in a simple geometry - superconducting
semi-infinitive sample $x\geq 0$. We assume that the external
magnetic field induction $B_e$ is parallel to the z-axis and the
magnetic field sweep rate $\dot{B_e}$ is constant. When the
magnetic field with the flux density $B_e$ is applied in the
direction of the z-axis, the transport current $j(x, t)$ and the
electric field $E(x, t)$ are induced inside the slab along the
y-axis. For this geometry the spatial and temporal evolution of
thermal and magnetic field perturbations
\begin{equation}
\begin{array}{l}
T=T_0+\Theta(x, t),\\
\quad\\
B=B_e+b(x, t),\\
\quad\\
V=V_0+v(x, t)\\
\end{array}
\end{equation}
are described by the following system of differential equations
[8, 11]

\begin{equation}
\dsf{d\Theta}{dt}=2v-\beta\Theta,
\end{equation}
\begin{equation}
\mu\dsf{dv}{dt}+v=-\dsf{db}{dx}+\beta\Theta,
\end{equation}
\begin{equation}
\dsf{db}{dt}=\left(\dsf{db}{dx}+b\right)+\left(\dsf{dv}{dx}+v\right),
\end{equation}
where $T_0(x)$, $B_e(x)$ and $V_0(x)$ are solutions to the
unperturbed equations, which can be obtained within a
quasi-stationary approximation. Here we have introduced the
following dimensionless variables
$$
b=\dsf{B}{B_e}=\dsf{B}{\mu_0j_cL},\quad
\Theta=\dsf{\nu\mu_0}{B_{e}^{2}}, \quad v=V\dsf{t_0}{L}.
$$
$$
z=\dsf{x}{L},\quad
\tau=\dsf{t}{t_0}=\dsf{\Phi_0}{\eta}\dsf{B_e}{\mu_0j_cL^2}t,
$$
and parameters
$$
\mu=\dsf{\Phi_0}{\mu_0\eta^2}\dsf{B_e}{L^2}m, \quad
\beta=\dsf{\mu_0j_{c}^{2}L^2}{\nu(T_c-T_0)}.
$$
Here $L=cB_e/\mu_0j_c$ is the magnetic field penetration depth,
which is determined from the following equation
\begin{equation}
B(x, t)=B_e+\mu_0j_c(x-L),
\end{equation}
with the appropriate boundary conditions
\begin{equation}
dB(0, t)=B_e,\quad B(L, t)=0.
\end{equation}

\vskip 0.5cm
\begin{center}
{\bf  3. Dispersion relation}
\end{center}
Assuming that the small thermal and magnetic perturbations has the
form $\Theta(x,t), b(x,t),  v(x,t)\sim\exp[\gamma t],$ where
$\gamma$ is the eigenvalue of the problem to be determined, we
obtain from equations (9)-(11) the following dispersion relations
to determine an eigenvalues of the problem

$$
(\gamma+\beta)
\dsf{d^2b}{dx^2}-[(\gamma+\beta)\mu-2\beta]\dsf{db}{dx}+[(\mu+1)\gamma^2+
$$
$$
+[(\mu-1)\beta- \mu-1]\gamma-(\mu-1)\beta]b=0
$$
The instability of the flux front is defined by the positive value
of the rate increase Re $\gamma$>0. An analysis of the dispersion
relation shows that, the grows rate is positive Re $\gamma$>0, if
$\mu>\mu_c=2$ and any small perturbations will grow with time. For
the case when $\mu<\mu_c$, the growth rate is negative and the
small perturbations will decay. At the critical value of
$\mu=\mu_c$, the growth rate is zero $\gamma$=0. For the specific
case, where $\mu=1$ the growth rate is determined by a stability
parameter $\beta$. Thus, the stability criterion can be written as

$$
\beta>1.
$$
For the case, where thermal effects is negligible ($\beta=1$) we
may obtain the following dispersion relation

\begin{equation}
\dsf{d^2b}{dx^2}-\mu\dsf{db}{dx}+(\gamma-1)(\mu+1)b=0.
\end{equation}
Seeking for $b\sim\exp(ikx)$ in dispersion relation, the growth
rate $\gamma$ dependence can be obtained as a functions of wave
number k.

\begin{center}
\includegraphics[width=3in]{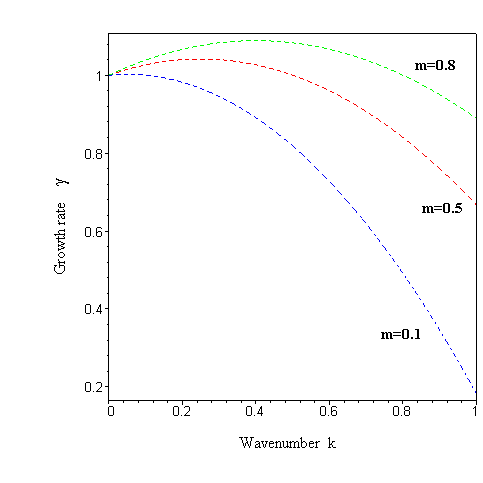}
\end{center}
\begin{center}
Fig.1.  The dependence of the growth rate on the wave number for
$\mu=0.1, 0.5, 0.8$.
\end{center}
We analyze the growth rate $\gamma$ of small perturbations as a
function of wave number k. When $k<k_c=\mu$ the growth rate is
positive and any small perturbations will grow with time. For wave
number $k>k_c$, the growth rate $\gamma$ is negative.
Consequently, the small perturbations always decay. It can be
shown that, for wave number $k=k_c$ the growth rate is zero
$\gamma=0$. As the wave number approaches zero $k\longrightarrow
0$ or infinity $k\longrightarrow \infty$ the growth rate
approaches $\gamma=1$ and small perturbations grow with time. As
the wave number approaches unity $k=1$ the growth rate is
determined by the value of $\mu$

$$
\gamma=\dsf{2\mu}{\mu+1}.
$$
For $\mu=0$ the growth rate is zero $\gamma=0$. For $\mu=1$ the
growth rate is unity $\gamma=1$. Since the growth rate is zero at
the critical wave number and approaches to unity in the limit of
zero wave number, there must exist a wave number in between that
maximizes the growth rate. Figs. (1-4) show the growing rate,
$\gamma$, as a function of the wave number k, for various values
of the vortex mass $\mu$. As the value of $\mu$ increases, the
corresponding growth rate increases.

\vskip 0.5cm
\begin{center}
{\bf  Conclusion}
\end{center}

Thus, in the present work we show that under some conditions flux
avalanche may occur in superconductor sample, which takes into
account the inertial properties of the vortices. It has been
noticed that a detailed theoretical study of this problem will be
presented in our further papers.

\vskip 0.5cm
\begin{center}
{\bf   Acknowledgements}
\end{center}

This study was supported by the NATO Reintegration Fellowship
Grant and Volkswagen Foundation Grant. Part of the computational
work herein was carried on in the Condensed Matter Physics at the
Abdus Salam International Centre for Theoretical Physics.

\vskip 0.5cm
\begin{center}
{\bf  References}
\end{center}

\begin{enumerate}
\item C. P. Bean, Phys. Rev. Lett., 8, 250 (1962); Rev. Mod.
Phys., 36, 31 (1964).

\item P. S. Swartz and S. P. Bean, J. Appl. Phys., 39, 4991
(1968).

\item S. L. Wipf, Cryogenics, 31, 936 (1961).

\item  R. G. Mints, and A. L. Rakhmanov, Rev. Mod. Phys., 53, 551
(1981).

\item R. G. Mints and A. L. Rakhmanov, Instabilities in
superconductors, Moscow, Nauka, 362 (1984).

\item V. V. Chabanenko, V. F. Rusakov, V. A. Yampol’skii, S.
Piechota, A. Nabialek, S. V. Vasiliev, and H. Szymczak,
cond-mat.0106379v2  (2002).

\item S. Vasiliev, A. Nabialek, V. Chabanenko, V. Rusakov, S.
Piechota, H. Szymczak, Acta Phys. Pol. A 109, 661 (2006).

\item A. Nabialek, S. Vasiliev, V. Chabanenko, V. Rusakov, S.
Piechota, H. Szymczak, Acta Phys. Pol. A, 114 (2008).

\item S. Vasiliev, A. Nabialek, V. F. Rusakov, L. V. Belevtsov,
V.V. Chabanenko and H. Szymczak,  Acta Phys. Pol. A, 118 (2010).

\item V. Rusakov, S. Vasilieva, V.V. Chabanenko , A. Yurov, A.
Nabialek, S. Piechotaa and H. Szymczak, Acta Phys. Pol. A, 109
(2006).

\item V. V. Chabanenko, V.F. Rusakov , A. Nabialek , S. Piechota ,
S. Vasiliev , H. Szymczak, Physica C, 369 (2002).

\item N. H. Zebouni, A. Venkataram, G. N. Rao, C. G. Grenier, J.
M. Reynolds, Phys. Rev. Lett., 13, 606 (1964).

\item H. Suhl, Phys. Rev. Lett., 14, 226 (1965).

\item H. T. Coffey, Cryogenics, 7, 73 (1967).

\item N. V. Kopnin. Pis’ma v ZhETF 27, 417 (1978).

\item G. Baym, E. Chandler. J. Low Temp. Phys., 50, 57  (1983).

\item E. B. Sonin, V. B. Geshkenbein, A. van Otterlo, G. Blatter.
Phys. Rev. B 57, 575 (1998).

\item M. J. Stephen, J. Bardin. Phys. Rev. Lett., 14 112 (1965).

\item G. E. Volovik, Pis’ma v ZhETF, 65, 201 (1997).

\item E. M. W. Coffey, Phys. Rev. B, 49, 9774 (1994).

\item J. I. Gittleman, B. Rosenblum. Journ. Appl. Phys., 39, 2617
(1968).

\newpage

\end{enumerate}
}\end{multicols}
\begin{center}
\includegraphics[width=3in]{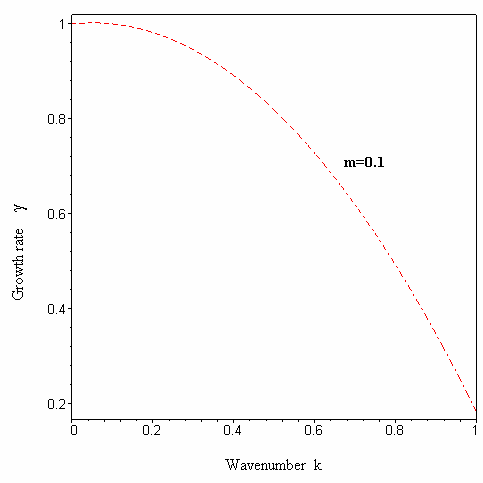}
\includegraphics[width=3in]{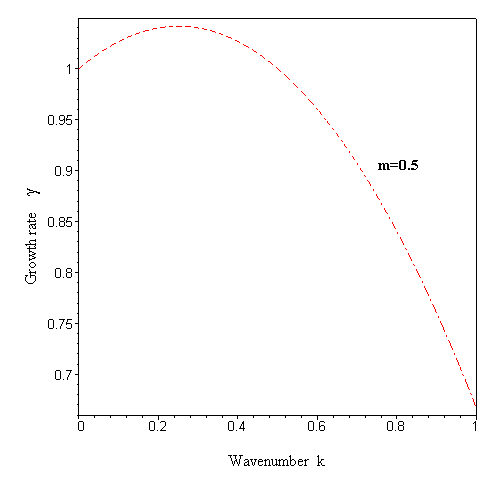}
\includegraphics[width=3in]{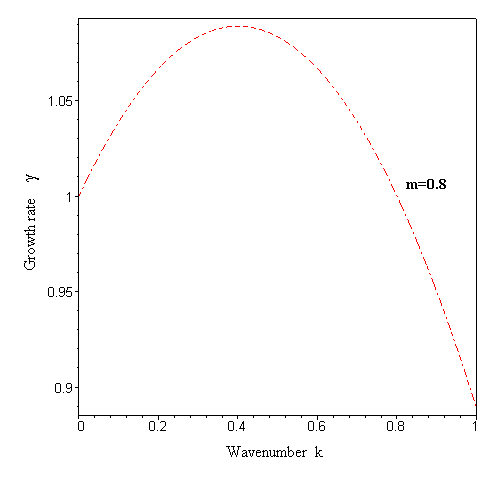}
\end{center}
\begin{center}
Fig.2-4.  The dependence of the growth rate on the wave number for
$\mu=0.1, 0.5, 0.8$.
\end{center}

\end{document}